\documentclass[11pt]{article}
\usepackage{placeins}
\usepackage{amsmath}
\usepackage{amsthm}
\usepackage{amssymb}
\usepackage{bigints}
\usepackage{graphicx}
\usepackage{keyval}
\usepackage{booktabs}
\usepackage{float}
\usepackage{setspace}
\usepackage{mathrsfs}
\usepackage{subfigure}
\usepackage{longtable}
\usepackage{verbatim}
\usepackage{arydshln}
\usepackage{enumerate}
\usepackage[usenames, dvipsnames]{color}
\usepackage{color}
\usepackage{footmisc}
\usepackage[authoryear]{natbib}
\usepackage{tikz}
\usepackage{soul}
\usepackage{multirow}
\usepackage{framed} 
\usepackage{color, colortbl}
\definecolor{LightCyan}{rgb}{0.88,1,1}
\definecolor{LightOrange}{rgb}{1,.84,.5}
\definecolor{LightGray}{gray}{0.9}
\usepackage{tablefootnote}
\usepackage{enumitem}
\usepackage{xcolor}
\usepackage{float}

\usepackage{pgfplots}
\pgfplotsset{compat=1.15,
             height=0.8\columnwidth,
             width=\columnwidth
             }\usepgfplotslibrary{ternary}

\usepackage{diagbox}
\usepackage{slashbox}
\usepackage{endnotes}
\usepackage{hyperref}
\hypersetup{
	pdfstartview = FitH,
	pdfauthor = {...},
	pdftitle = {...},
	pdfkeywords = {...; ...; ...; ...},
	colorlinks = true,
	linkcolor = orange,
	urlcolor = black,
	citecolor = blue,
	linktocpage=true}
	
\usepackage{tabularx}
\usepackage{lscape}    
\usepackage{booktabs}  
\usepackage{array}     

\allowdisplaybreaks

\usepackage{geometry}
\graphicspath{{Plots/}}
\numberwithin{equation}{section}

\newtheorem*{proposition*}{Proposition}

\numberwithin{assumption}{section}

\theoremstyle{definition}

\begin{document}
\begin{titlepage}
\pagenumbering{gobble}
\title{\LARGE \textbf{Chapter: Delegation and Lobbying}}
\author{\textbf{Thomas Groll}\thanks{School of International and Public Affairs, Columbia University, New York, NY; tgroll@columbia.edu.} \\ Columbia University \\ \and \textbf{Sharyn O'Halloran}\thanks{School of International and Public Affairs and Department of Political Science, Columbia University, New York, NY, USA; Department of Economics and Political Science, Trinity College Dublin, Ireland; so33@columbia.edu.} \\ Columbia University \\ Trinity College Dublin}

\date{October 3, 2025}
\maketitle

\begin{center}
\textbf{Prepared for \textit{``The Handbook of Strategic Analysis of Politics''} \\ edited by Olga Shvetsova}
\bigskip
\bigskip
\bigskip
\end{center}

\begin{abstract}
This chapter examines the link between delegation and lobbying, two themes central to political economy. Delegation models explore how legislatures manage uncertainty and control bureaucratic agents, while lobbying models analyze how organized interests influence policy through contributions, information, and advocacy. We review the growing body of research that integrates these literatures, showing how the prospect of lobbying affects legislative incentives to delegate and how the structure of delegated authority shapes lobbying strategies. We highlight common-agency frameworks that capture the recursive relationship between delegation and lobbying and empirical studies documenting how venue choice, information provision, and interest group mobilization mediate delegation outcomes. We also review applications to agency oversight and fiscal policy. Finally, we present a model of regulatory rule-making that embeds lobbying directly into the delegation decision, offering predictions for both theory and empirical analysis.
\end{abstract}

\par\vspace{2ex}

\vfill

\end{titlepage}
	\setlength{\parindent}{.25in}

\doublespacing

\newpage
\pagenumbering{arabic}	
	\setcounter{page}{1}		

\section{Introduction}

Delegation and lobbying are central themes in the political economy literature. Delegation models have long been studied to manage uncertainty and oversee bureaucratic agents \citep{fiorina1982,epsteinhalloran1994,epsteinhalloran1999,hubershipan2002,volden2002jleo}. Lobbying is viewed as a conduit through which organized interests shape policy—whether by supplying campaign contributions, transmitting information, or exerting direct pressure during administrative rule-making \citep{stigler1971, peltzman1975,becker1983,grossmanhelpman1994}. Each area of this literature is extensive and, until recently, has developed in parallel.

A small but growing body of research indicates that the two phenomena are closely connected \citep{epsteinhalloran1995,sloof2000, DeBievre2005,bennedsenfeldmann2006sje,boehmkegailmardpatty2006}. Legislators’ willingness to delegate authority to regulatory agencies depends on how likely interest groups are to mobilize. The strategies of interest groups also depend on how delegated authority is structured. For example, delegating regulatory power to regulators creates new opportunities for lobbying. At the same time, the potential for lobbying influences whether and how authority is initially delegated. This recursive relationship guides our analysis.

The purpose of this chapter is to unify this diverse literature within a single framework. We focus on two areas of delegation where lobbying plays a vital role: fiscal policy (taxation and spending) and agency rule-making authority. For the first area, we review key contributions; for the second, we develop a model that incorporates lobbying directly into the delegation process. Our goal is to clarify when delegation and lobbying support each other, when they oppose each other, and what our integrated approach means for both theory and empirical research.

\section{Literature Review}

The classic dilemma identified by \citet{fiorina1982} is why a legislature intent on shaping policy would delegate authority to an agency instead of legislating directly. This question motivates two related but distinct areas of research: what motivates legislators to delegate, and how they design the institutions that accompany the delegated authority. Forty years of research make clear that these two are tightly linked. The choice between legal versus administrative policymaking, and between rules versus discretion, depends heavily on the level of uncertainty and conflict at both the legislative and regulatory stages.

\subsection{Delegation}

Recent research views delegation as a form of risk management. Legislatures delegate less during times of high partisan conflict, such as under divided government, but are more willing to delegate when agencies can provide expertise that legislatures lack \citep{epsteinhalloran1994, epsteinhalloran1999, volden2002ajps,hubershipan2002,wiseman2009}. Once delegated, discretionary authority is rarely taken back, leading to the gradual expansion of the regulatory state \citep{volden2002jleo}. Technical contributions refine these insights by identifying when discontinuous or interval forms of delegation are optimal, particularly when principals cannot credibly commit to ex post actions \citep{melumadshibano1991,alonsomatuschek2008,gailmard2009}. Others emphasize how institutional rules---such as supermajority requirements, appointment procedures, or bureaucratic career incentives---reduce drift and define the scope of delegation \citep{bendormeirowitz2004, gailmardpatty2007, callenderkrehbiel2014}. 

Two themes emerge. First, delegation is often designed to balance the risks of uncertainty against the costs of agency drift. Second, delegation is constrained by partisan conflict, although external interest group pressures can facilitate it. For example, demands for policy change from interest groups or the public can decrease uncertainty and make delegation more appealing.

\subsection{Interest Group Lobbying}

The literature on delegation explores how legislators can control policymaking under conditions of uncertainty. Studies on lobbying examine the challenges legislators face in exerting influence under asymmetric power or information. Early political economy models characterize lobbying as regulatory capture, where agencies protect industries in exchange for their political support. Political outcomes reflect the balance of marginal votes gained versus those lost \citep{stigler1971, peltzman1975, becker1983}. This ``market for regulation'' approach underscores agencies' susceptibility to organized interests but offers limited insight into the specific mechanisms used to influence regulators.

Subsequent research views lobbying as a strategic exchange of resources and information. Menu-auction models \citep{bernheim1986menu,grossmanhelpman1994} formalize the interest group contributions as contingent transfers to politicians, as campaign contributions to support favorable candidates \citep{baron1994, grossmanhelpman1996}, or both \citep{fellimerlo2007}. This contrasts with informational models where interest groups transmit expertise to policymakers whose own knowledge is limited \citep{ crawfordsobel1982, milgromroberts1986, gilligankrehbiel1989, pottersvandwinden1992, austensmithwright1992, hopenhaynlohmann1996}. Policymakers may not only lack expertise but also be time- and resource-constrained in gathering or assessing policy-relevant information or implementing policies. Interest group competition for limited access to policymakers can then result in financial contributions with the goal to transmit information \citep{austensmith1995,lohmann1995,cotton2009,cotton2012}, agenda distortions \citep{cottondellis2016,dellisoak2019}, or de facto subsidies of financial or informational resources to relax politicians' resource constraints strategically \citep{halldeardorff2006,ellisgroll2020}. Recent studies analyze the conditions under which agencies become partially ``captured,'' not because interest groups buy them outright but because eliciting industry information requires aligning agency preferences more closely with those industries \citep{gailmardpatty2019}. Similarly, interest groups consider legislators' preferences for policies in legislative informational lobbying \citep{bennedsenfeldmann2002,schnakenberg2015,schnakenberg2017,awad2020,awadminaudier2024} and try to identify their optimal lobbying targets among legislators \citep{grollprummer2016,dellis2023}. There has been also more attention to lobbyists as intermediaries -- acting between interest groups and policymakers who may either be former staffers of policymakers or policymakers themselves, participating in the ``revolving-door'', or working for commercial lobbying firms -- who offer contacts and personal connections \citep{BlanesiVidal2012,Bertrand2014} or policy-relevant resources \citep{grollellis2014,grollellis2017,ellisgroll2025}.\endnote{For a review of the literature on lobbyists as intermediaries in the lobbying process and the the various principal-agent problems that may arise, see \citet{ellisgroll2024}.} \citet{Figueiredo2014}, \citet{Bombardini2020}, and \citet{garlicketal2025} offer exhaustive reviews of the recent empirical literature in economics and political science, and \citet{schnakenbergturner2024} on the more recent theoretical literature in lobbying.

Despite these advances, few studies directly examine lobbying within the context of agency rule-making, where discretion and industry expertise are most closely linked. The more agencies rely on industry participants' input and expert advice, either because of a lack of expertise or resources, the more likely they are to \textit{de facto} ``re-delegate'' or transfer rule-making authority to regulated industries. 

This knowledge gap underscores the need to integrate the literature on delegation and lobbying. By analyzing the interplay between the institutional design of delegated authority and interest group pressure, we can gain a deeper understanding of whether agencies deviate from legislative intent and how organized interests strategically influence this deviation through electoral contributions and administrative pressure. Yet delegation models that ignore lobbying miss a key piece of the puzzle: legislators delegate not only under uncertainty about agency expertise but also in anticipation of how interest groups will respond. This insight motivates the recent literature that explicitly links lobbying and delegation.

\subsection{Delegation and Lobbying}

The most direct link between the delegation and lobbying literature occurs in models that consider them jointly determined. This research covers both fiscal policy and regulatory rule-making. Still, the core logic remains the same: legislators anticipate lobbying when deciding whether and how to delegate, while interest groups adapt their strategies based on the scope of authority granted.

\subsubsection*{Oversight}
Early work on delegation emphasizes how interest groups provide information that shapes oversight of agencies’ regulatory authority. For example, classic theories of ``fire-alarm'' versus ``police patrol'' oversight \citep{mccubbinsschwartz1984} and administrative procedure as political control \citep{McCubbinsNollWeingast1987, McCubbinsNollWeingast1989} show how organized interests serve as monitors in delegated systems. Formal theoretical work deepened this logic: \citet{epsteinhalloran1995} show how lobbying can discipline agencies by reducing informational asymmetries, while \citet{DixitGrossmanHelpman1997} introduce a general common-agency framework for competing lobbies. \citet{bendoretal2001}; and \citet{bendormeirowitz2004} further show how spatial delegation models interact with bias and uncertainty. Together, these strands establish the informational and strategic foundations for subsequent delegation--lobbying models.

\subsubsection*{Public Finance}
Several studies apply these concepts to fiscal policymaking. \citet{DeBievre2005} analyze trade policy in the United States and European Union, showing empirically that delegation and legislative control vary with the share of tradable goods. \citet{mazzavanwinden2008} extend theory with a hierarchical model in which legislatures may delegate fiscal authority to bureaucrats while interest groups lobby at multiple levels. Their analysis shows that multi-stage lobbying can, under certain conditions, mitigate rather than worsen capture. \citet{Sorge2010} emphasizes ``lobbying-consistent delegation,'' in which legislatures design delegation rules with expected lobbying in mind. \citet{limaetal2017} highlight the informational role of lobbying in centralized versus decentralized fiscal decision-making: centralization may improve welfare when it ensures informed lobbying reaches the relevant decision-maker, though it can also increase the risks of capture.

\subsubsection*{Rule-Making and Common Agency}
The regulatory environment explicitly combines delegation and lobbying. An important contribution is \citet{sloof2000}, who models interest groups’ decision to lobby politicians rather than bureaucrats as a common agency game. In this model, lobbying is not only an ex post effort to influence delegated policy but also an ex ante factor determining whether delegation takes place. Politicians might rationally delegate to a biased bureaucrat because doing so can lead to more informative lobbying, even if it causes some agency drift.

This common-agency approach has led to many extensions. \citet{bennedsenfeldmann2006sje} explore lobbying during the implementation phase of bureaucracy, showing how it influences legislative incentives to delegate. \citet{mazzavanwinden2008} hierarchical model broadens Sloof’s idea to multiple levels of government. \citet{evansetal2008} along with \citet{britopereiravareda2013}, demonstrate that delegating to biased regulators can help address time-inconsistency issues in regulation, supporting Sloof’s finding that moderate bias can sometimes improve welfare. \citet{sorge2015} includes bureaucratic appointments, illustrating how principals consider lobbying when choosing agency heads. 

Empirical research supports these theoretical insights. \citet{boehmkegailmardpatty2006} show that interest groups strategically select lobbying venues based on how authority is delegated. \citet{You2017Ex} documents that groups engage in ex post lobbying during rule-making after the enactment of legislation. \citep{yackee2006, YackeeYackee2009} reveal that lobbying influences the content and scope of administrative rulemaking, and \citet{ballawright2001} examine how advisory committees provide organized interests with access to agencies. Large-scale studies such as \citet{baumgartneretal2009} underscore the difficulty of policy change despite intense lobbying. Comparative studies, such as those by \citet{rasmussenreher2019}, \citet{hanegraaffberkhout2019}, and \citet{gigerkluver2016}, reveal how venue choice and information sharing vary across political systems. \citet{gailmardpatty2013} review this literature, emphasizing how agency expertise interacts with interest group information within delegated authority.

Overall, this body of work demonstrates that lobbying and delegation are endogenous. Legislators design delegation rules considering lobbying pressures, and interest groups leverage delegation structures to enhance their influence. The common-agency framework captures this recursive relationship, essential to understanding modern policymaking functions.

The literature thus reveals two key insights. First, a legislature's decision to delegate authority cannot be examined in isolation: the incentives of interest groups are an integral part of the delegation calculus. Second, lobbying is not merely a distortion layered on top of delegation; its form and effectiveness depend on the structure of delegated authority. In the next section, we build on these insights by developing a formal model incorporating lobbying directly into the delegation process. The model shows how contributions and administrative pressure interact with congressional decisions on the status quo policy and delegated discretion, producing hypotheses that directly address ongoing debates about agency capture, legislative control, and the strategic influence of interest groups.

\section{Delegation and Lobbying in Rule-Making}

\cite{epsteinhalloran1995} develop a formal model of congressional oversight in which interest groups play a strategic role. In their framework, lobbying operates as an informational signal that reduces asymmetries between Congress and agencies: groups can “sound the alarm” to discipline bureaucrats, thereby improving legislative control. This early integration of delegation and lobbying highlights how oversight institutions and interest group mobilization jointly determine agency behavior. \cite{sloof2000} builds on this insight by formalizing the decision of interest groups to lobby politicians or bureaucrats as a common agency game. Whereas Epstein and O’Halloran emphasize lobbying as an informational oversight mechanism, Sloof shows that the anticipation of lobbying itself can shape the delegation decision, with legislators rationally delegating to biased bureaucrats precisely because it yields more informative lobbying equilibria. We next extend this work by embedding lobbying directly into the delegation process, demonstrating how contributions and administrative pressure influence congressional choices regarding status quo policy and delegated discretion.

\subsection{A Model}

Our model examines strategic interactions among Congress (C), an executive agency (A), and an interest group (I) representing a regulated industry. As in the standard delegation model, all actors have ideal points and quadratic preferences over policy outcomes in a one-dimensional outcome space: $u_i(x)=-(x-x_i)^2$ for $x\in X=\mathbb{R}$ and $i=C,A,I$. Without loss of generality, we assume that $x_I=0$ and $x_A>0$. We further assume that $x_C>0$, so if the value of $x$ represents the strength of regulation, the industry prefers less regulation than either Congress or the agency.

Congress may delegate authority to make policy decisions to the agency, and final policy outcomes are a function of both the policy $p$ chosen and an external shock $\omega$ according to the equation $x=p+\omega$. The external shock $\omega$ is uniformly distributed: $\omega \sim U\in[-R,R]$. When delegating, Congress can also place discretionary constraints on the agency's policy choice. Thus, Congress can set a status quo policy $p_0$ and a discretion limit $d$ so that $|p-p_0|\leq d$.

Where we differ from the standard model is our assumption that the interest group can affect outcomes directly by lobbying Congress and the agency.\footnote{For a model of administrative lobbying in rule-making, see \citet{grolletal2019chapter}.} At the legislative stage, the interest group can offer policy-contingent contributions to Congress in exchange for legislation that sets the status quo policy and delegates discretion. We consider a menu-auction approach \citep{bernheim1986menu, grossmanhelpman1994}. The interest group's financial contributions, $m=M(p_0,d)$, to Congress are weighted as benefits from policies, $0<\lambda<1$, and financial contributions, $1-\lambda$. At the rule-making stage, the interest group can also exert lobbying pressure on the agency after it announces its proposed rule of $p_A$. In particular, we assume that $p=p_A-e$, where $e$ is the amount of costly effort exerted by the interest group. We envision this effort coming in the form of presenting analyses and testimony at the notice and comment stage of rule-making, broader lobbying efforts aimed at legislators, executive officials, and the public at large to weaken industry regulations, or legal threats and litigation. The cost to the group of this effort is $c(e)$, with $c'>0$ and $c''>0$.

For the sake of concreteness, we take $c(e)=\alpha e^{2}$, so that $\alpha>0$ measures the relative cost of lobbying to the interest group, and low values of $\alpha$ indicate the ability to exert greater pressure on regulators. This lobbying also reduces the agency's utility by an amount $-\beta e$, where $\beta\geq0$ is the cost to the agency of having its original proposals moved back towards the interest group's ideal point. It is possible to set $\beta=0$, so that $\beta>0$ indicates that the agency would prefer to implement a given policy outcome directly, rather than propose a stricter regulation and have the industry lobby to weaken the agency's proposal. 

\begin{figure}
\centering
\includegraphics{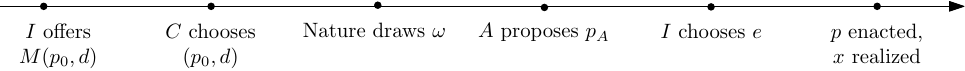}
\caption{\label{fig:timeline2}Model with Legislative and Administrative Lobbying.}
\end{figure}

Overall, then, $u_C=-\lambda(x-x_C)^2+(1-\lambda)M(p_0,d)$, $u_I=-x^2-\alpha e^2-M(p_0,d)$, and $u_A=-(x-x_A)^2-\beta e$, where $x=p_A-e+\omega$. The order of events is illustrated in Figure \ref{fig:timeline2} and is the following: First, the interest group offers policy-contingent contributions, $M(p_0,d)$. Second, Congress observes the interest group's lobbying and sets the status quo and the discretion limit, $(p_0, d)$. Then nature draws $\omega$, which is observed by both the agency and the interest group. Fourth, the agency proposes its policy rule $p_A$. Fifth, the interest group observes $p_A$ and chooses its lobbying effort level $e$. Finally, policy $p$ is enacted, and policy outcome $x$ with corresponding utility levels is realized. We solve the game for its subgame-perfect Bayesian-Nash equilibrium.

\subsection{Interest Group's Administrative Lobbying}

Starting at the end of the game and working backward, for a given policy proposal $p_A$ and shock $\omega$, the industry will set its lobbying effort $e$ to maximize $-(p_A-e+\omega)^2-\alpha e^2$. This leads to lobbying in the amount of
\begin{equation}\label{model_IG_best-response-lobbying}
e^\ast(p_A)=\frac{p_A+\omega}{1+\alpha}.
\end{equation}
Thus, positive amounts of lobbying are exerted whenever $p_A+\omega>0$, and it goes to zero when the agency accommodates the industry by making final policy outcomes equal to the interest group's ideal point. Note that $\partial e^\ast/\partial p_A>0$, so that the interest group spends fewer resources lobbying an agency with preferences closer to their own. Further, the greater the shock, $\partial e^\ast/\partial \omega >0$, the more lobbying effort the interest group undertakes.

\subsection{Agency's Policy Choices}

Knowing the interest group's best response to the announced policy rule, $e^\ast(p_A)$, the agency will propose policy rule $p_A$ to maximize $-[p_A-e^\ast(p_A)+\omega-x_A]^2-\beta e^\ast(p_A)$, yielding 
\begin{equation}\label{model_Agency_policy-response}
p_A^\ast=\frac{(1+\alpha)(2x_A\alpha-\beta)}{2\alpha^2} -\omega
\end{equation}
iff $|p_A^\ast - p_0| \leq d$ and constrained by $(p_0, d)$, otherwise.

Combining equations (\ref{model_IG_best-response-lobbying}) and (\ref{model_Agency_policy-response}), final policy outcomes will be:
\begin{equation}\label{policy_outcome_nolobbying}
\tilde{x}=p_A^\ast-e^\ast(p_A^\ast)+\omega=x_A-\frac{\beta}{2\alpha}.
\end{equation}
Notice that this point lies in the interval between the agency's and the interest group's ideal points as long as $x_A>\beta/2\alpha$. We refer to the term $\beta/2\alpha$ as the \textit{effective} administrative lobbying pressure, which is a combination of the agency's lobbying burden and the industry's lobbying cost. For values of $\beta$ greater than or equal to $2\alpha x_A$, the agency sets $p_A$ such that $x=0$ and the industry does not lobby. Consequently, if the burden to the agency of industry lobbying is high relative to the industry's lobbying cost, the agency may propose a policy rule so that the industry gets its ideal point of no regulation without the interest group having to actively lobby to obtain this outcome.\footnote{This can serve as a convenient definition of agency capture: the mere threat of lobbying causes the agency to accommodate industry wishes so that in equilibrium the industry escapes effective government control without actually having to expend resources to do so.} However, for greater ideal points of policy outcomes, $x_A>\beta/2\alpha$, the industry undertakes a lobbying effort, and the effective administrative lobbying pressure is not sufficient to prevent regulation.

\subsection{Congress's Policy Choice}

\begin{figure}
\centering
\includegraphics{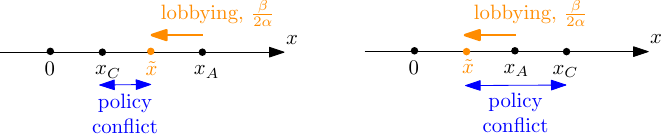}
\caption{\label{fig:ideal-pressure}Ideal Points, Lobbying Pressure, and Policy Outcomes.}
\end{figure}

Congress, on the other hand, would like the agency to set policy so that the outcome, net of industry lobbying, is Congress's ideal point: $p_A-e^\ast(p_A)+\omega=x_C$, which simplifies to $p_C^\ast=x_C\frac{(1+\alpha)}{\alpha}-\omega$. For any given value of $\omega$, then, Congress's and the agency's ideal policies differ by an amount of 
\begin{equation}\label{conflict_gap}
p_C^\ast-p_A^\ast=\left(\frac{1+\alpha}{\alpha}\right)\left(x_A-x_C-\frac{\beta}{2\alpha}\right).
\end{equation}
This expression goes to zero when 
\begin{equation}\label{ally_principle_condition}
 x_A = x_C + \frac{\beta}{2\alpha}.
\end{equation}
Thus, the ``ally principle'' fails to hold in our model: Congress prefers an agency not with its own ideal point, but one biased slightly against the industry, since policy outcomes are a convex combination of the agency's ideal point and the industry's desire for no regulation and lobbying pressure. In other words, lobbying by the interest group mitigates the preference conflict between Congress and the agency if $x_C<x_A$, as illustrated on the left in Figure \ref{fig:ideal-pressure}, or increases the conflict if $x_C>x_A$, as illustrated on the right, over policy outcomes $\tilde{x}$.

However, Congress is unable to observe the policy shock $\omega$ and can only determine the status quo policy and delegate discretion to the agency to affect policy outcomes. Furthermore, Congress observed the interest group's offer of financial contribution contingent on legislation, $M(p_0,d)$.

\subsubsection{Status Quo and Discretion}

Congress, anticipating the agency's policy rule proposal and the interest group's pressure -- as well as a resulting policy conflict -- can set its optimal status quo policy $p_0$ and discretion limit $d$. The policy outcomes are given that any status quo, discretion, and external shock are
\begin{equation}
x^\ast = \begin{cases}
p_0 + d + \omega & \text{ if } -R \leq \omega < \tilde{x} - p_0 - d \\
\tilde{x} & \text{ if } \tilde{x} - p_0 - d \leq \omega \leq  \tilde{x} - p_0 + d \\
p_0 - d + \omega & \text{ if } \tilde{x} - p_0 + d < \omega \leq R .
\end{cases}
\end{equation}

As it is common for the menu-auction approach \citep{bernheim1986menu}, we are solving for a jointly efficient solution that maximizes the weighted sum of expected utilities of Congress and the interest group with the relative weights of Congress' policy focus, $\lambda$, and contribution focus, $(1-\lambda)$. The optimal legislation $(p_0^L,d^L)$ maximizes then
\begin{equation}
\lambda EU_C(p_0,d) + (1-\lambda)EU_I(p_0,d).
\end{equation} 
Congress's expected utility follows from 
\begin{eqnarray}\label{expect_utility_Congress_nolobbying}
EU_C(p_0,d) &=& -\int_{-R}^{\tilde{x}-d-p_0} \frac{(p_0+d+\omega-x_C)^2}{2 R}  d\omega - \int_{\tilde{x}-d-p_0}^{\tilde{x}+d-p_0} \frac{(\tilde{x}-x_C)^2}{2 R} d\omega  \\ \nonumber
&& -\int_{\tilde{x}+d-p_0}^R \frac{(p_0-d+\omega -x_C)^2}{2 R}  d\omega
\end{eqnarray}
and the interest group's expected utility can be described by 
\begin{eqnarray}\label{expect_utility_Industry_nolobbying}
EU_I(p_0,d) &=& -\int_{-R}^{\tilde{x}-d-p_0} \frac{(p_0+d+\omega)^2}{2 R}  d\omega - \frac{1+\alpha}{\alpha}\int_{\tilde{x}-d-p_0}^{\tilde{x}+d-p_0} \frac{\tilde{x}^2}{2 R} d\omega  \\ \nonumber
&& -\int_{\tilde{x}+d-p_0}^R \frac{(p_0-d+\omega)^2}{2 R}  d\omega
\end{eqnarray}
with $\tilde{x}=x_A-\frac{\beta}{2\alpha}$ as described in (\ref{policy_outcome_nolobbying}) when the agency is expected to propose $p_A^\ast$ and the lobby exerts administrative lobbying effort $e^\ast(p_A)$. The optimal status quo policy follows then from
\begin{equation}\label{optimal_status_quo_lobbying}
\frac{\partial ...}{\partial p_0} = \frac{2(d-R)(p_0-\lambda x_C)}{R}=0 \text{ } \Rightarrow \text{ } p_0^L=\lambda x_C,
\end{equation}
In other words, the optimal status quo policy is identical to Congress's ideal point weighted by its policy focus; and when Congress's desire for contribution increases, lower $\lambda$, then the status quo policy decreases and moves closer to the industry's preferred policy. This also implies that if Congress did not value policy outcomes at all, $\lambda=0$, then it would be captured by the interest group and prefer no regulation in exchange for small contributions, as it would not be expensive for the interest group to induce $p_0^L=0$. If Congress did not value contributions at all, $\lambda=1$, then $p_0$ would be chosen as Congress's ideal point $x_C$.

Further, the optimal discretion limit follows then from
\begin{equation}
\frac{\partial ...}{\partial d} = \frac{2\alpha\lambda x_C\tilde{x}- (1+\alpha-\lambda)\tilde{x}^2 +\alpha\left((d-R)^2 + p_0(p_0-2\lambda x_C)\right)}{\alpha R}=0 
\end{equation}
with $p_0^L=\lambda x_C$ such that
\begin{equation}
(d^L-R)^2 = \left(\frac{1+\alpha-\lambda}{\alpha}\right)\tilde{x}^2 + \lambda\left(\lambda x_C^2 + 2x_C\tilde{x}\right).
\end{equation}

\begin{figure}
\centering
\subfigure[$\alpha=\beta=1,\lambda=.5$]{\label{fig:equal}\includegraphics[width=2in]{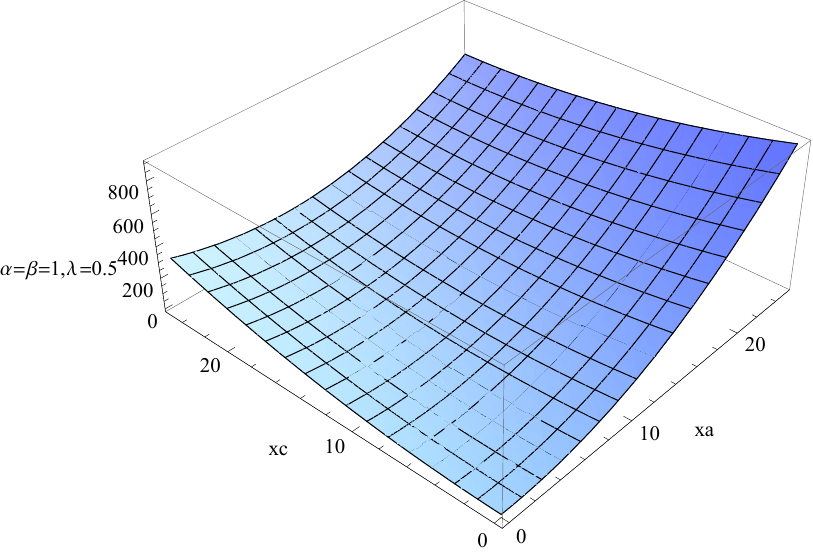}}
\subfigure[$\alpha=1,\beta=5,\lambda=.5$]{\label{fig:greaterbeta}\includegraphics[width=2in]{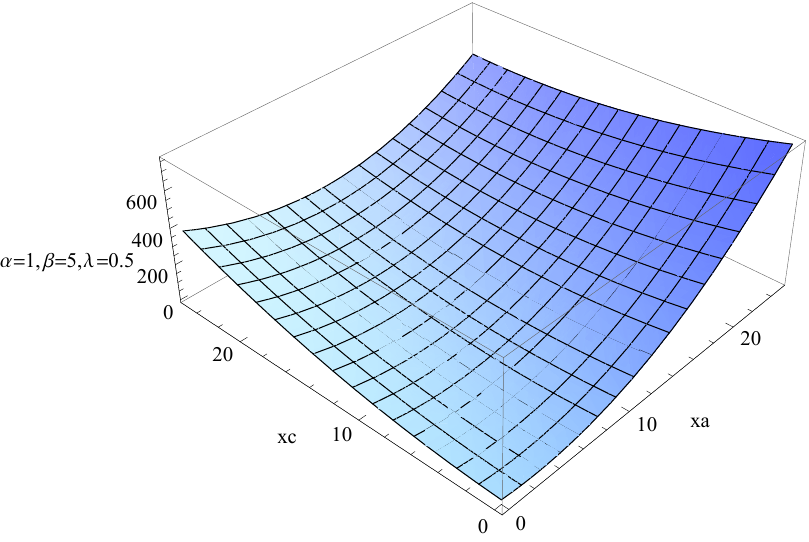}}
\subfigure[$\alpha=5,\beta=1,\lambda=.5$]{\label{fig:greateralpha}\includegraphics[width=2in]{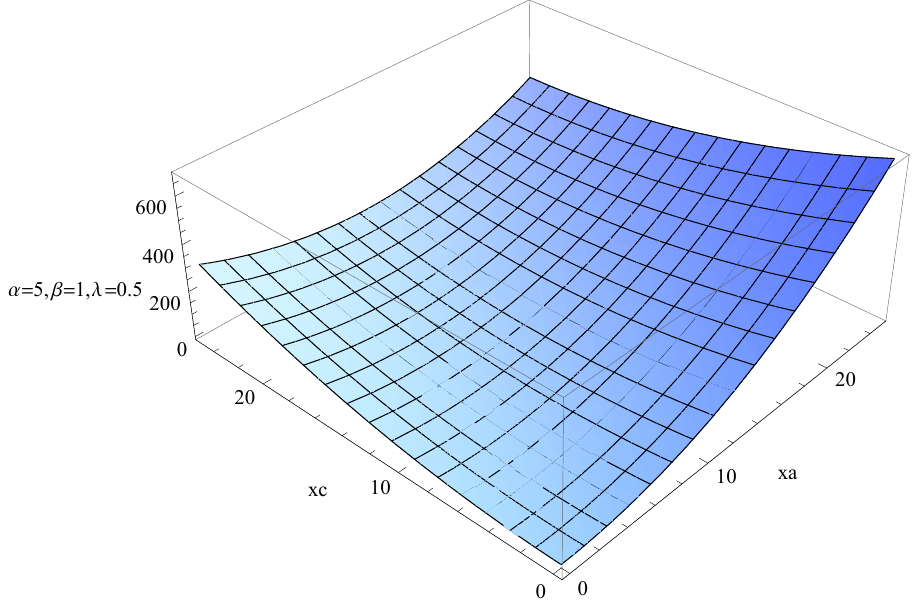}}
\caption{Preference Conflict and Delegated Discretion with $(d^L-R)^2$.}
\label{fig:preferencesdelegation}
\end{figure}

 In Figure \ref{fig:preferencesdelegation} we illustrate the delegation of discretion in more general terms and plot $(d^L-R)^2$ as a function of $x_A$ and $x_C$ for given parameter values for $\alpha$, $\beta$, and $\lambda$. The function implies an inverse relationship -- meaning that low values of $d^L$ imply a greater value of $(d^L-R)^2$ given $d^L<R$. When Congress values policy outcomes and contributions similarly, intermediate values of $\lambda$, the delegated discretion is greatest when Congress' and the agency's ideal points are closer to industry interests ($x_C$ and $x_A$ closer to 0) and the agency is less biased than Congress (to the left of the diagonal of $x_C=x_A$ because of $\lambda<1$). As Congress becomes more biased, $x_C$ increases, and the agency is industry-friendlier than Congress, it reduces discretion. 
 
For an alternative illustration, we consider the two extreme cases of
\begin{eqnarray}\label{delegated_discretion_piecewise}
d^L = \begin{cases}
R - \frac{1+\alpha}{\alpha}\left|x_A - \frac{\beta}{2\alpha}\right|  & \text{ if } \lambda = 0 \\
R - \left|x_A - x_C - \frac{\beta}{2\alpha}\right| = d^\ast          & \text{ if } \lambda = 1 \\
\end{cases} .
\end{eqnarray}

If Congress is purely policy-motivated, then discretion is greater the smaller the policy conflict between Congress and the pressured agency, the agency is slightly more biased than Congress (``ally principle does not hold''), agency's lobbying burden is high and administrative lobbying mitigates the policy conflict between policymakers, the industry's lobbying ability is high and administrative lobbying amplifies the policy conflict between policymakers. However, suppose Congress is not motivated by policies. In that case, discretion increases the interest group's effectiveness in influencing the agency's rule-making towards the industry's ideal point and Congress's status quo policy. Note that in all three considered scenarios, greater policy uncertainty and greater $R$ imply greater discretion.

\subsection{Summary of Hypotheses from Model}

We can summarize our analysis with the following testable hypotheses. Table \ref{ComparativeStatics} relates each of the nine hypotheses to the model parameters and denotes the equations from which they are derived.  

\begin{itemize}
\item \textbf{Hypothesis 1: Best-response lobbying} Lobbying efforts are decreasing as (i) the agency's proposed policy becomes closer to the industry's ideal point; (ii) relative lobbying costs increase; (iii) the magnitude of policy shocks decreases.\\  
\textit{\textbf{Logic:} Lobbying is strongest when agencies propose tougher rules, costs are low, and shocks are large.}  

\item \textbf{Hypothesis 2: Agency policy choice} The agency chooses a lower policy level in response to greater external shocks and lobbying pressure by the industry, but rises with the agency’s own ideal point.\\  
\textit{\textbf{Logic:} Agencies back off tough policies when shocks or lobbying rise, but tougher agencies still push higher.}  

\item \textbf{Hypothesis 3: Industry's lobbying} The industry's lobbying effort is (i) decreasing in the agency's lobbying burden; (ii) increasing in the agency's ideal point; (iii) decreasing in the industry's lobbying cost if the agency's ideal point is greater than twice the lobbying pressure, and vice versa.\\  
\textit{\textbf{Logic:} Industries lobby more when agencies are tougher and lobbying is cheap, but back off when regulators already bear heavy costs.}  

\item \textbf{Hypothesis 4: Lobbying outcome} If the lobbying cost to the agency is high relative to industry cost, then a mere threat of lobbying induces the agency to set policies preferred by the industry.\\  
\textit{\textbf{Logic:} Cheap lobbying and costly resistance lead agencies to accommodate industry without actual pressure.}  

\item \textbf{Hypothesis 5: Conflict gap} The policy conflict between Congress and the agency is increasing in the agency's lobbying burden but decreasing in the lobby's relative cost if Congress prefers a higher policy level. Hence, lobbying can amplify or mitigate conflict.\\  
\textit{\textbf{Logic:} Heavy lobbying burdens widen Congress–agency gaps, but costly lobbying narrows them.}  

\item \textbf{Hypothesis 6: Ally principle} The ``ally principle'' does not hold, and Congress prefers an agency that is slightly more biased against industry (as resistance against lobbying pressure).\\  
\textit{\textbf{Logic:} Congress wants regulators tougher than itself to offset lobbying.}  

\item \textbf{Hypothesis 7: Optimal status quo } Congress's status quo policy in the presence of contributions from the SIG is (i) increasing in Congress's ideal point and (ii) increasing in Congress's policy motivation but decreasing as it would value SIGs' contributions more.\\  
\textit{\textbf{Logic:} The status quo follows Congress’s ideal when policy dominates, but shifts toward industry when money matters.}  

\item \textbf{Hypothesis 8: Delegated discretion} Congress delegates greater discretion when (i) there is more policy uncertainty for Congress; (ii) Congress's and the agency's ideal points are closer to industry interests; (iii) Congress values contributions but is more biased than the agency; (iv) Congress does not value contributions and (iv') the agency is slightly more biased; (iv'') administrative lobbying pressure mitigates the preference conflict between Congress and the agency.\\  
\textit{\textbf{Logic:} Discretion expands with uncertainty and alignment, and contracts when bias or lobbying sharpen conflict.}  

\item \textbf{Hypothesis 9: Policy outcome} Regulatory policy outcomes are higher when (i) industry's lobbying cost is high; (ii) agency's lobbying burden is low; (iii) agency's ideal point is high; (iv) Congress' status quo policy is higher (higher policy motivation and ideal point).\\  
\textit{\textbf{Logic:} Regulation is strongest when lobbying is costly, agencies are tough, and Congress sets a high baseline.}  
\end{itemize}

\section{Conclusion and Future Research}

Delegation and lobbying are endogenous: legislators delegate with lobbying in mind, and interest groups lobby in response to delegated authority. This recursive relationship shapes oversight, fiscal policy, and rule-making. Our analysis shows that when lobbying eases conflict, deepens it, or when the mere threat of mobilization bends agencies toward industry. Future work should examine these dynamics in repeated and multi-level settings and use new data to test when lobbying informs policy or distorts it through capture. Integrating lobbying into delegation theory sharpens predictions and highlights how institutional design channels the influence of organized interests.

\theendnotes

\newpage

\begingroup
		\setlength{\bibsep}{10pt}
    \setstretch{.75}
\bibliographystyle{agsm}
\bibliography{ChapterXX_Groll-OHalloran_Delegation-Lobbying}

@article{austensmithwright1992,
title = "Competitive lobbying for a legislator's vote",
author = "David Austen-Smith and Wright, \{John R.\}",
year = "1992",
month = jul,
doi = "10.1007/BF00192880",
language = "English (US)",
volume = "9",
pages = "229--257",
journal = "Social Choice and Welfare",
issn = "0176-1714",
publisher = "Springer New York",
number = "3",
}

@Article{bennedsenfeldmann2006sje,
  author    = {Bennedsen, M. and Feldmann, S.},
  title     = {Lobbying Bureaucrats},
  journal   = {Scandinavian Journal of Economics},
  year      = {2006},
  volume    = {108},
  number    = {4},
  pages     = {643-668},
}

@Article{bennedsenfeldmann2002,
  author    = {Bennedsen, Morten and Feldmann, Sven},
  title     = {Lobbying Legislatures},
  journal   = {Journal of Political Economy},
  year      = {2002},
  volume    = {110},
  number    = {4},
  pages     = {919-946},
}

@Article{bernheim1986menu,
  author        = {Bernheim, B.D. and Whinston, M.D.},
  title         = {Menu Auctions, Resource Allocation, and Economic Influence},
  journal       = {The Quarterly Journal of Economics},
  year          = {1986},
  volume        = {101},
  number        = {1},
  pages         = {1-31},
  publisher     = {Oxford University Press},
}

@Article{grossmanhelpman1994,
  author    = {Grossman, Gene and Helpman, Elahan},
  title     = {Protection for Sale},
  journal   = {American Economic Review},
  year      = {1994},
  volume    = {84},
  number    = {4},
  pages     = {833-850},
}

@Book{epsteinhalloran1999,
  title     = {Delegating Powers: A Transaction Cost Politics Approach to Policy Making under Separate Powers},
  publisher = {Cambridge University Press},
  year      = {1999},
  author    = {Epstein, David and O'Halloran, Sharyn},
  address   = {New York, NY},
}

@Article{epsteinhalloran1995,
  author  = {Epstein, David and O'Halloran, Sharyn},
  title   = {A Theory of Strategic Oversight: Congress, Lobbyists, and the Bureaucracy},
  journal = {Journal of Law, Economics, and Organization},
  year    = {1995},
  volume  = {11},
  pages   = {227-255},
}

@Article{epsteinhalloran1994,
  author  = {Epstein, David and O'Halloran, Sharyn},
  title   = {Administrative Procedures, Information and Agency Discretion},
  journal = {American Journal of Political Science},
  year    = {1994},
  volume  = {38},
  pages   = {697-722},
}

@Article{mazzavanwinden2008,
  author  = {Mazza, Isidoro and van Winden, Frans},
  title   = {An Endogenous Policy Model of Hierarchical Government},
  journal = {European Economic Review},
  year    = {2008},
  volume  = {52},
  pages   = {133-149},
}

@Article{bendormeirowitz2004,
  author  = {Bendor, Jonathon and Meirowitz, Adam},
  title   = {Spatial Model of Delegation},
  journal = {American Political Science Review},
  year    = {2004},
  volume  = {98},
  number  = {2},
  pages   = {293-310},
}

@Article{gailmardpatty2007,
  author  = {Gailmard, Sean and Patty, John W.},
  journal = {American Journal of Political Science},
  title   = {Slackers and Zealots: Civil Service, Policy Discretion, and Bureaucratic Expertise},
  year    = {2007},
  number  = {4},
  pages   = {873-889},
  volume  = {51},
}

@Article{alonsomatuschek2008,
  author  = {Alonso, Ricardo and Matouschek, Niko},
  title   = {Optimal Delegation},
  journal = {Review of Economic Studies},
  year    = {2008},
  volume  = {75},
  pages   = {259-293},
}

@Article{volden2002ajps,
  author  = {Volden, Craig},
  title   = {A Formal Model of the Politics of Delegation in a Separation of Powers System},
  journal = {American Journal of Political Science},
  year    = {2002},
  volume  = {46},
  number  = {1},
  pages   = {111-133},
}

@Article{stigler1971,
  author  = {Stigler, George J.},
  title   = {The Theory of Economic Regulation},
  journal = {Bell Journal of Economics},
  year    = {1971},
  volume  = {2},
  pages   = {3-21},
}

@Article{wiseman2009,
  author  = {Wiseman, Alan E.},
  title   = {Delegation and Positive-Sum Bureaucracies},
  journal = {Journal of Politics},
  year    = {2009},
  volume  = {71},
  number  = {3},
  pages   = {998-1014},
}

@Book{hubershipan2002,
  title     = {Deliberate Discretion? The Institutional Foundations of Bureaucratic Autonomy},
  publisher = {Cambridge University Press},
  year      = {2002},
  author    = {Huber, John D. and Shipan, Charles R.},
  address   = {New York, NY},
}

@Article{boehmkegailmardpatty2006,
  author  = {Boehmke, Frederick J. and Gailmard, Sean and Patty, John Wiggs},
  title   = {Whose Ear to Bend? Information Sources and Venue Choice in Policy Making},
  journal = {Quarterly Journal of Political Science},
  year    = {2006},
  volume  = {1},
  number  = {2},
  pages   = {139-169},
}

@Article{gailmard2009,
  author  = {Gailmard, Sean},
  title   = {Discretion Rather than Rules: Choice of Instruments to Control Bureaucratic Policy Making},
  journal = {Political Analysis},
  year    = {2009},
  volume  = {17},
  number  = {1},
  pages   = {25-44},
}

@Article{melumadshibano1991,
  author  = {Melumad, Nahum D. and Shibano, Toshiyuki},
  title   = {Communication in Settings with No Transfers},
  journal = {RAND Journal of Economics},
  year    = {1991},
  volume  = {22},
  number  = {2},
  pages   = {173-198},
}

@Article{peltzman1975,
  author  = {Peltzman , Sam},
  title   = {The Effects of Automobile Safety Regulation},
  journal = {Journal of Political Economy},
  year    = {1975},
  volume  = {83},
  number  = {4},
  pages   = {677-726},
}

@Article{becker1983,
  author  = {Becker, Gary S.},
  title   = {A Theory of Competition Among Pressure Groups for Political Influence},
  journal = {Quarterly Journal of Economics},
  year    = {1983},
  volume  = {98},
  number  = {3},
  pages   = {371-400},
}

@Article{volden2002jleo,
  author  = {Volden, Craig},
  title   = {Delegating Power to Bureaucracies: Evidence from the States},
  journal = {Journal of Law, Economics, and Organization},
  year    = {2002},
  volume  = {18},
  number  = {1},
  pages   = {187-220},
}

@Article{mccubbinsschwartz1984,
  author  = {McCubbins, Mathew D. and Schwartz, Thomas},
  title   = {Congressional Oversight Overlooked: Police Patrols versus Fire Alarms},
  journal = {American Journal of Political Science},
  year    = {1984},
  volume  = {28},
  number  = {1},
  pages   = {165-179},
}

@Article{hopenhaynlohmann1996,
  author  = {Hopenhayn, Hugo and Lohmann, Susanne},
  title   = {Fire-Alarm Signals and the Political Oversight of Regulatory Agencies},
  journal = {Journal of Law, Economics, and Organization},
  year    = {1996},
  volume  = {12},
  number  = {1},
  pages   = {196-213},
}

@Article{gailmardpatty2019,
  author  = {Gailmard, Sean and Patty, John Wiggs},
  title   = {Giving Advice vs. Making Decisions: Transparency, Information, and Delegation},
  journal = {Political Science Research and Methods},
  year    = {2019},
  volume  = {7},
  number  = {3},
  pages   = {471 - 488},
}

@Article{callenderkrehbiel2014,
  author  = {Callender, Steven and Krehbiel, Keith},
  title   = {Gridlock and Delegation in a Changing World},
  journal = {American Journal of Political Science},
  year    = {2014},
  volume  = {58},
  number  = {4},
  pages   = {819-834},
}

@Article{fiorina1982,
  author  = {Morris P. Fiorina},
  title   = {Legislative choice of regulatory forms: Legal process or administrative process?},
  journal = {Public Choice},
  year    = {1982},
  volume  = {39},
  pages   = {33-66},
}

@Article{evansetal2008,
  author   = {Joanne Evans and Paul Levine and Francesc Trillas},
  journal  = {International Journal of Industrial Organization},
  title    = {Lobbies, delegation and the under-investment problem in regulation},
  year     = {2008},
  issn     = {0167-7187},
  number   = {1},
  pages    = {17-40},
  volume   = {26},
}

@Article{sorge2010,
  author  = {Sorge, Marco},
  journal = {Economics Bulletin},
  title   = {Lobbying-consistent Delegation and Sequential Policy Making},
  year    = {2010},
  month   = {01},
  pages   = {3088-3102},
  volume  = {30},
  issue   = {4},
}

@article{sorge2015,
author = {Sorge, Marco},
title = {Lobbying (strategically appointed) bureaucrats},
journal = {Constitutional Political Economy},
year = {2015},
pages = {171-189},
volume = {26},
}

@Book{gailmardpatty2013,
  author    = {Sean Gailmard and John W. Patty},
  publisher = {Chicago University Press},
  title     = {Learning While Governing},
  year      = {2013},
}

@InBook{grolletal2019chapter,
  author    = {Thomas Groll and Sharyn O'Halloran and Geraldine McAllister},
  chapter   = {Trends and Delegation in U.S. Financial Market Regulation},
  editor    = {Sharyn O'Halloran and Thomas Groll},
  pages     = {57-81},
  publisher = {Columbia University Press},
  title     = {After the Crash: Financial Crises and Regulatory Responses},
  year      = {2019},
}

@Article{bendoretal2001,
  author  = {John Bendor and Amihai Glazer and Thomas Hammond},
  journal = {Annual Review of Political Science},
  title   = {Theories of Delegation},
  year    = {2001},
  pages   = {235-269},
  volume  = {4},
}

@Article{You2017Ex,
  author    = {You, Hye Young},
  journal   = {The Journal of Politics},
  title     = {Ex Post, Lobbying},
  year      = {2017},
  issn      = {0022-3816},
  month     = {10},
  number    = {4},
  pages     = {1162--1176},
  volume    = {79},
  publisher = {University of Chicago Press},
}

@article{Sloof2000,
  author = {Sloof, Randolph},
  title = {Interest Group Lobbying and the Delegation of Policy Authority},
  year = {2000},
  volume = {12},
  number = {3},
  pages = {247--274},
  journaltitle = {Economics \& Politics},
  date = {2000}
}

@article{DeBievre2005,
  author = {De Bièvre, Dirk and Dür, Andreas},
  title = {Constituency Interests and Delegation in European and American Trade Policy},
  year = {2005},
  volume = {38},
  number = {10},
  pages = {1271--1296},
  journaltitle = {Comparative Political Studies},
  date = {2005}
}

@Article{limaetal2017,
  author       = {Lima, Francisco and Moreira, Humberto and Verdier, Thierry},
  journal      = {American Economic Journal: Microeconomics},
  title        = {Centralized Decision Making and Informed Lobbying},
  year         = {2017},
  number       = {4},
  pages        = {324-355},
  volume       = {9},
  date         = {2017},
  journaltitle = {American Economic Journal: Microeconomics},
}

@article{McCubbinsNollWeingast1987,
  author = {McCubbins, Matthew D. and Noll, Roger G. and Weingast, Barry R.},
  title = {Administrative Procedures as Instruments of Political Control},
  year = {1987},
  volume = {3},
  number = {2},
  pages = {243--277},
  journaltitle = {Journal of Law, Economics, and Organization},
  date = {1987}
}

@article{McCubbinsNollWeingast1989,
  author = {McCubbins, Matthew D. and Noll, Roger G. and Weingast, Barry R.},
  title = {Structure and Process, Politics and Policy: Administrative Arrangements and the Political Control of Agencies},
  year = {1989},
  volume = {75},
  number = {2},
  pages = {431--482},
  journaltitle = {Virginia Law Review},
  date = {1989}
}

@article{DixitGrossmanHelpman1997,
  author = {Dixit, Avinash and Grossman, Gene M. and Helpman, Elhanan},
  title = {Common Agency and Coordination: General Theory and Application to Government Policy Making},
  year = {1997},
  volume = {105},
  number = {4},
  pages = {752--769},
  journaltitle = {Journal of Political Economy},
  date = {1997}
}

@Article{gigerkluver2016,
  author       = {Giger, Nathalie and Klüver, Heike},
  title        = {Voting Against Your Constituents? How Lobbying Affects Representation},
  year         = {2016},
  number       = {1},
  pages        = {190--205},
  volume       = {60},
  date         = {2016},
  journaltitle = {American Journal of Political Science},
}

@Article{ballawright2001,
  author  = {Steven J. Balla and John R. Wright},
  journal = {American Journal of Political Science},
  title   = {Interest Groups, Advisory Committees, and Congressional Control of the Bureaucracy},
  year    = {2001},
  number  = {4},
  pages   = {799-812},
  volume  = {45},
}

@Article{yackee2006,
  author  = {Jason Webb Yackee and Susan Webb Yackee},
  journal = {Journal of Politics},
  title   = {A Bias Towards Business? Assessing Interest Group Influence on the U.S. Bureaucracy},
  year    = {2006},
  number  = {1},
  pages   = {128-139},
  volume  = {68},
}

@Article{YackeeYackee2009,
  author   = {Yackee, Jason Webb and Yackee, Susan Webb},
  journal  = {Regulation \& Governance},
  title    = {Divided government and US federal rulemaking},
  year     = {2009},
  number   = {2},
  pages    = {128-144},
  volume   = {3},
}

@Article{rasmussenreher2019,
  author  = {Anne Rasmussen and Stefanie Reher},
  journal = {Comparative Political Studies},
  title   = {Civil Society Engagement and Policy Representation in Europe},
  year    = {2019},
  number  = {11},
  pages   = {1648-1676},
  volume  = {52},
}

@Article{hanegraaffberkhout2019,
  author  = {Marcel Hanegraaf and Joost Berkhout},
  journal = {Journal of European Public Policy},
  title   = {More Business as Usual? Explaining Business Bias across Issues and Institutions in the European Union},
  year    = {2019},
  number  = {6},
  pages   = {843-862},
  volume  = {26},
}

@Book{baumgartneretal2009,
  author    = {Frank R. Baumgartner and Jeffrey M. Berry and Marie Hojnacki and Beth L. Leech and David C. Kimball},
  publisher = {University of Chicago Press},
  title     = {Lobbying and Policy Change: Who Wins, Who Loses, and Why},
  year      = {2009},
}

@Article{britopereiravareda2013,
  author  = {Duarte Brito and Pedro Pereira and Joao Vareda},
  journal = {The B.E. Journal of Economic Analysis \& Policy},
  title   = {Investment, Dynamic Consistency and the Sectoral Regulator's Objective},
  year    = {2013},
  number  = {2},
  volume  = {13},
}

@Article{AustenSmith1995,
  author    = {Austen-Smith, D.},
  journal   = {American Political Science Review},
  title     = {Campaign Contributions and Access},
  year      = {1995},
  number    = {3},
  pages     = {566-581},
  volume    = {89},
}

@Article{Cotton2012,
  author    = {Cotton, Christopher},
  journal   = {Journal of Public Economics},
  title     = {Pay-to-Play Politics: Informational Lobbying and Contribution Limits when Money Buys Access},
  year      = {2012},
  pages     = {369-386},
  volume    = {96},
}

@Article{Cotton2009,
  author    = {Cotton, Christopher},
  journal   = {Journal of Public Economics},
  title     = {Should We Tax or Cap Political Contributions? A Lobbying Model with Policy Favors and Access},
  year      = {2009},
  pages     = {831-842},
  volume    = {93},
}

@Article{cottondellis2016,
  author  = {Cotton, Christopher and Dellis, Arnaud},
  journal = {Journal of Law, Economics, and Organization},
  title   = {Informational Lobbying and Agenda Distortion},
  year    = {2016},
  number  = {4},
  volume  = {32},
}

@Article{crawfordsobel1982,
  author    = {Crawford, Vincent P. and Sobel, Joel},
  journal   = {Econometrica},
  title     = {Strategic Information Transmission},
  year      = {1982},
  number    = {6},
  pages     = {1431-1451},
  volume    = {50},
}

@Article{Figueiredo2014,
  author    = {de Figueiredo, J. M. and Richter, Brian K.},
  journal   = {Annual Review of Political Science},
  title     = {Advancing the Empirical Research on Lobbying},
  year      = {2014},
  pages     = {163-185},
  volume    = {17},
}

@InBook{ellisgroll2024,
  author    = {Christopher J. Ellis and Thomas Groll},
  chapter   = {Commercial Lobbying Firms: Lobbying as Business},
  editor    = {Karsten Mause and Andreas Polk},
  pages     = {221â€“245},
  publisher = {Springer "Studies in Public Choice"},
  title     = {The Political Economy of Lobbying: Channels of Influence and their Regulation},
  year      = {2024},
  volume    = {43},
}

@Article{ellisgroll2020,
  author  = {Ellis, Christopher J. and Thomas Groll},
  journal = {American Political Science Review},
  title   = {Strategic Legislative Subsidies: Informational Lobbying and the Cost of Policy},
  year    = {2020},
  month   = feb,
  number  = {1},
  pages   = {179-205},
  volume  = {114},
}

@Article{gilligankrehbiel1989,
  author  = {Thomas W. Gilligan and Krehbiel, Keith},
  journal = {American Journal of Political Science},
  title   = {Asymmetric Information and Legislative Rules with a Heterogeneous Committee},
  year    = {1989},
  number  = {2},
  pages   = {459-490},
  volume  = {33},
}

@Article{halldeardorff2006,
  author    = {Hall, Richard L. and Deardorff, Alan V.},
  journal   = {American Political Science Review},
  title     = {Lobbying as Legislative Subsidy},
  year      = {2006},
  number    = {1},
  pages     = {69-84},
  volume    = {100},
}

@Article{Lohmann1995,
  author    = {Lohmann, Susanne},
  journal   = {Public Choice},
  title     = {Information, Access, and Contributions: A Signaling Model of Lobbying},
  year      = {1995},
  number    = {3-4},
  pages     = {267-284},
  volume    = {85},
}

@Article{milgromroberts1986,
  author    = {Milgrom, Paul and Roberts, John},
  journal   = {RAND Journal of Economics},
  title     = {Relying on the Information of Interested Parties},
  year      = {1986},
  number    = {1},
  pages     = {18-32},
  volume    = {17},
}

@Article{pottersvandwinden1992,
  author    = {Potters, Jan and van Winden, Frans},
  journal   = {Public Choice},
  title     = {Lobbying and Asymmetric Information},
  year      = {1992},
  pages     = {269-292},
  volume    = {74},
}

@Article{schnakenbergturner2024,
  author  = {Keith E. Schnakenberg and Ian R. Turner},
  journal = {Annual Review of Political Science},
  title   = {Formal Theories of Special Interest Influence},
  year    = {2024},
  pages   = {401-421},
  volume  = {27},
}

@Article{grollellis2017,
  author    = {Groll, Thomas and Christopher J. Ellis},
  journal   = {Economic Inquiry},
  title     = {Repeated Lobbying by Commercial Lobbyists and Special Interests},
  year      = {2017},
  number    = {4},
  pages     = {1868-1897},
  volume    = {55},
}

@Article{grollellis2014,
  author    = {Groll, Thomas and Ellis, Christopher J.},
  journal   = {European Economic Review},
  title     = {A Simple Model of the Commercial Lobbying Industry},
  year      = {2014},
  pages     = {299-316},
  volume    = {70},
}

@Article{schnakenberg2017,
  author  = {Schnakenberg, Keith E.},
  journal = {American Journal of Political Science},
  title   = {Informational Lobbying and Legislative Voting},
  year    = {2017},
  number  = {1},
  pages   = {129-145},
  volume  = {61},
}

@Article{fellimerlo2007,
  author  = {Leonardo Felli and Antonio Merlo},
  journal = {Journal of the European Economic Association},
  title   = {If You Cannot Get Your Friends Elected, Lobby Your Enemies},
  year    = {2007},
  number  = {2-3},
  pages   = {624-635},
  volume  = {5},
}

@Article{baron1994,
  author  = {David Baron},
  journal = {American Political Science Review},
  title   = {Electoral Competition with Informed and Uninformed Voters},
  year    = {1994},
  number  = {1},
  pages   = {33-47},
  volume  = {88},
}

@Article{grossmanhelpman1996,
  author  = {Gene M. Grossman and Elhanan Helpman},
  journal = {Review of Economic Studies},
  title   = {Electoral Competition and Special Interest Politics},
  year    = {1996},
  number  = {2},
  pages   = {265-286},
  volume  = {63},
}

@Article{dellisoak2019,
  author  = {Arnaud Dellis and Mandar Oak},
  journal = {Journal of Law, Economics, and Organization},
  title   = {Informational Lobbying and Pareto-Improving Agenda Constraint},
  year    = {2019},
  number  = {3},
  pages   = {579-618},
  volume  = {35},
}

@Article{dellis2023,
  author  = {Arnaud Dellis},
  journal = {Journal of Economic Theory},
  title   = {Legislative Informational Lobbying},
  year    = {2023},
  volume  = {208},
}

@Article{awad2020,
  author  = {Emiel Awad},
  journal = {American Journal of Political Science},
  title   = {Persuasive Lobbying with Allied Legislators},
  year    = {2020},
  number  = {4},
  pages   = {938-951},
  volume  = {64},
}

@Article{awadminaudier2024,
  author  = {Emiel Awad and Clement Minaudier},
  journal = {American Journal of Political Science},
  title   = {Friendly Lobbying under Time Pressure},
  year    = {2024},
  number  = {2},
  pages   = {529-543},
  volume  = {68},
}

@Article{grollprummer2016,
  author  = {Thomas Groll and Anja Prummer},
  journal = {mimeo},
  title   = {Whom to Lobby? Targeting in Political Networks},
  year    = {2016},
}

@Article{schnakenberg2015,
  author  = {Schnakenberg, Keith E.},
  journal = {Journal of Economic Theory},
  title   = {Expert Advice to a Voting Body},
  year    = {2015},
  pages   = {102-113},
  volume  = {160},
}

@Article{Bertrand2014,
  author    = {Bertrand, Marianne and Bombardini, Matilde and Trebbi, Francesco},
  journal   = {American Economic Review},
  title     = {Is It Whom You Know or What You Know? An Empirical Assessment of the Lobbying Process},
  year      = {2014},
  number    = {12},
  pages     = {3885-3920},
  volume    = {104},
}

@Article{BlanesiVidal2012,
  author    = {Blanes i Vidal, J. and Draca, M. and Fons-Rosen, C.},
  journal   = {American Economic Review},
  title     = {Revolving Door Lobbyists},
  year      = {2012},
  number    = {7},
  pages     = {3731-3748},
  volume    = {102},
}

@Article{Bombardini2020,
  author  = {Matilde Bombardini and Francesco Trebbi},
  journal = {Annual Review of Economics},
  title   = {Empirical Models of Lobbying},
  year    = {2020},
  month   = aug,
  pages   = {391-413},
  volume  = {12},
}

@Article{ellisgroll2025,
  author  = {Christopher J. Ellis and Thomas Groll},
  journal = {mimeo},
  title   = {Who lobbies whom? Special interests and hired guns},
  year    = {2025},
}

@Article{garlicketal2025,
  author  = {Alex Garlick and Wiebke Marie Junk and Heath Brown},
  journal = {Annual Review of Political Science},
  title   = {How Lobbying Matters},
  year    = {2025},
  pages   = {457-475},
  volume  = {28},
}
\endgroup

\newpage
\begin{landscape}
\begin{table}[htbp]
\centering
\renewcommand{\arraystretch}{1.2}
\setlength{\tabcolsep}{5pt}

\begin{tabularx}{\linewidth}{
  >{\hsize=.01\hsize}X
  >{\hsize=.18\hsize}X
  >{\hsize=.12\hsize}X
  >{\hsize=.28\hsize}X
  >{\hsize=.40\hsize}X}
\toprule
\multicolumn{1}{c}{\textbf{Prop.}} &
\multicolumn{1}{c}{\textbf{Label}} &
\multicolumn{1}{c}{\textbf{Equation Ref.}} &
\multicolumn{1}{c}{\textbf{Statement}} &
\multicolumn{1}{c}{\textbf{Comparative statics}} \\
\midrule

H1 & Best-response lobbying & Eq.~(\ref{model_IG_best-response-lobbying}) &
Lobbying falls with proximity, higher costs, or smaller shocks. &
$\partial e^*/\partial p_A>0$; $\partial e^*/\partial \omega>0$; $\partial e^*/\partial \alpha<0$; $e^*>0$ iff $p_A+\omega>0$. \\

H2 & Agency policy choice & Eq.~(\ref{model_Agency_policy-response}) &
Agency lowers policy with larger shocks, higher burden, or lower ideal point. &
$\partial p_A^*/\partial \omega<0$; $\partial p_A^*/\partial \beta<0$; $\partial p_A^*/\partial x_A>0$. \\

H3 & Industry lobbying (via $p_A^*$) & Eqs.~(\ref{model_IG_best-response-lobbying}), (\ref{model_Agency_policy-response}) &
Lobbying falls with agency burden, rises with agency ideal point, falls with group cost. &
$\partial e^*/\partial \beta<0$; $\partial e^*/\partial x_A>0$; $\partial e^*/\partial \alpha<0$. \\

H4 & Lobbying outcome & Eq.~(\ref{policy_outcome_nolobbying}) &
Threat of lobbying alone can yield industry’s outcome. &
If $\beta \ge 2\alpha x_A \Rightarrow \tilde{x}=0$ and $e^*=0$. \\

H5 & Conflict gap & Eq.~(\ref{conflict_gap}) &
Congress–agency gap varies with costs and preferences. &
If $x_C<x_A$: gap $\downarrow$ in $\beta$, $\uparrow$ in $\alpha$; if $x_C>x_A$: gap $\uparrow$ in $\beta$, $\downarrow$ in $\alpha$. \\

H6 & Ally principle & Eq.~(\ref{ally_principle_condition}) &
Congress prefers slightly tougher agency (Ally principle fails). &
Preferred $x_A=x_C+\tfrac{\beta}{2\alpha}$; bias $\uparrow$ in $\beta$, $\downarrow$ in $\alpha$. \\

H7 & Optimal status quo & Eq.~(\ref{optimal_status_quo_lobbying}) &
Status quo reflects Congress’s ideal point and weight on policy vs. money. &
$p_0^L=\lambda x_C$; $\uparrow$ in $x_C$; $\downarrow$ as $\lambda\downarrow$. \\

H8 & Delegated discretion & Eq.~(\ref{delegated_discretion_piecewise}) &
Discretion grows with uncertainty, falls with conflict and burden. &
$d^L\uparrow R$; $d^L\uparrow \alpha$; $d^L\downarrow \beta$; $d^L\downarrow$ in $|x_A|$ if $\lambda=0$, $d^L\downarrow$ in $|x_A-x_C|$ if $\lambda=1$. \\

H9 & Policy outcome & Eq.~(\ref{policy_outcome_nolobbying}) &
Outcomes rise with agency ideal point, fall with agency burden, rise with lobbying cost. &
$\partial \tilde{x}/\partial x_A>0$; $\partial \tilde{x}/\partial \beta<0$; $\partial \tilde{x}/\partial \alpha>0$. Higher $p_0$ raises outcomes when constrained. \\
\bottomrule
\end{tabularx}
\caption{Hypotheses and comparative statics from the model, with equation references and descriptive labels.}
\label{ComparativeStatics}
\end{table}
\end{landscape}

\end{document}